\documentclass[%
 reprint,
 amsmath,amssymb,
 aps,
]{revtex4-2}

\usepackage{graphicx}
\usepackage{dcolumn}
\usepackage{bm}
\usepackage{bbold}
\usepackage{multirow}
\usepackage{caption}
\usepackage{bbding}
\usepackage{pifont}
\usepackage{wasysym}
\usepackage{amssymb}
\usepackage{xcolor}
\usepackage{tikz-feynman} 

\begin{document}

\preprint{APS/123-QED}

\title{Alternative 3-3-1 models with exotic electric charges}

\author{Eduard Suarez}
\email{esuarezar@gmail.com }
\affiliation{Departamento de F\'\i sica, Universidad de Nari\~no, Calle 18 Carrera 50,
A.A. 1175,  San Juan de Pasto, Colombia}
\author{Richard H. Benavides}
\email{richardbenavides@itm.edu.co}
\affiliation{ Facultad de Ciencias Exactas y Aplicadas, Instituto Tecnol\'ogico Metropolitano,
Calle 73 No 76 A - 354 , V\'ia el Volador, Medell\'in, Colombia}
\author{Yithsbey Giraldo}
\email{yithsbey@gmail.com}
\affiliation{Departamento de F\'\i sica, Universidad de Nari\~no, Calle 18 Carrera 50,
A.A. 1175,   San Juan de Pasto, Colombia}
\author{William A. Ponce}
\email{william.ponce@udea.edu.co}
\affiliation{Instituto de F\'isica, Universidad de Antioquia, Calle 70 No.~52-21, Apartado A\'ereo 1226, Medell\'in, Colombia}
\author{Eduardo Rojas}
\email{eduro4000@gmail.com}
\affiliation{Departamento de F\'\i sica, Universidad de Nari\~no, Calle 18 Carrera 50,
A.A. 1175,   San Juan de Pasto, Colombia}




\date{\today}

\begin{abstract}

We report the most general classification of 3-3-1 models with $\beta =\sqrt{3}$. We found several solutions where anomaly cancellation occurs among fermions of different families. These solutions are particularly interesting as they generate non-universal heavy neutral vector bosons. Non-universality in the SM fermion charges under an additional  gauge group generates Charged Lepton Flavor Violation (CLFV) and Flavor Changing Neutral Currents~(FCNC); we discuss under what conditions the new models can evade constraints coming from these processes. In addition, we also report the Large Hadron Collider~(LHC) constraints.

\end{abstract}

\maketitle

\section{Introduction}
Models with exotic fermions   based on the gauge group symmetry $SU(3)\otimes SU(3)\otimes U(1)$ (hereafter 3-3-1 models for short) have been proposed since the early 1970s~\cite{Schechter:1973nqg,Albright:1974nd,Fayet:1974fj,Fritzsch:1976dq,Segre:1976rc,Mohapatra:1976nv,Lee:1977qs,Lee:1977tx,Langacker:1977ae,Georgi:1978bv,Singer:1980sw};  
however, many of these models lacked important properties of what is known nowadays as 3-3-1 models. For a model to be interesting from a modern perspective~\cite{Pleitez:2021abk}, it must be chiral, the triangle anomalies must be canceled out only with a number of generations multiple of 3, and most importantly, it must contain the Standard Model~(SM).

In the 1990s, non-universal models without exotic leptons gained {popularity} as they were very convenient in addressing  flavor problems~\cite{Pisano:1992bxx,Frampton:1992wt}.
These models have also been helpful in explaining neutrino masses~\cite{Queiroz:2010rj,Caetano:2012qc,Ferreira:2011hm,Mohapatra:1979ia,Dias:2005yh,Dias:2010vt,Cogollo:2009yi,Ky:2005yq,Palcu:2006ti,Dias:2011sq}, dark matter~\cite{Fregolente:2002nx,Long:2003hht,Filippi:2005mt,Mizukoshi:2010ky,Profumo:2013sca,Kelso:2013nwa,RodriguesdaSilva:2014gbi,Queiroz:2013lca,Cogollo:2014jia,Kelso:2014qka,Dong:2014esa}, charge quantization~\cite{deSousaPires:1998jc}, strong CP violation~\cite{Pal:1994ba,Dias:2003iq}, muon anomalous magnetic moment ($g\text{-}2$ muon 
anomaly)~\cite{deJesus:2020ngn,Hue:2020wnn,Hue:2021zyw} and flavor anomalies~\cite{Buras:2012dp,Addazi:2022frt,Descotes-Genon:2017ptp,Wei:2017ago}.

Pleitez and Frampton proposed the non-universal 3-3-1 models~\cite{Pisano:1992bxx,Frampton:1992wt} as examples of electroweak extensions with lepton number violation, where the number of families is determined by anomaly cancellation.
In the literature, there are many examples of models without exotic electric charges, 
these models have been appropriately classified, and their phenomenology is well known~\cite{Ponce:2001jn,Ponce:2002sg,Benavides:2021pqx}.  
The original model of Pleitez and Framton has exotic electric charges in the quark sector and  corresponds to what is known in the literature as $\beta =\sqrt{3}$~\cite{Pleitez:2021abk}. As far as we know, an exhaustive classification of models with this $\beta $ does not exist in the literature, and therefore a work in this line is  {necessary}. 
It is important to notice that there are solutions for arbitrary $\beta $~\cite{Diaz:2004fs}; however, this solution does not account for  all the possible models for a given $\beta $.
As we will see, the parameter $\beta$  cannot be arbitrarily large, from the matching conditions $|\beta|\lessapprox \cot \theta_W\sim 1.8 $. 
This condition constitutes a very important restriction regarding the possible realizations of the 3-3-1 symmetry at low energies as it limits the number of possible non-trivial  cases to a countable set.

In section~\ref{sec:sec2}, we review the basics of the 3-3-1 models. In section~\ref{sec:3}, we propose sets of fermions corresponding to families of quarks and leptons with the left-handed triplets, anti-triplets, and singlets of $SU(3)_L$. In section~\ref{sec:4}, we show the anomaly-free sets~(AFSs) that constitute the basis for model building. This section lists all possible 3-3-1 models with $\beta=\sqrt{3}$ modulo lepton vector arrays. Finally, in section~\ref{sec:5}, we show the collider constraints and the conditions the models must satisfy to avoid FCNC and CLFV restrictions. 
\section{\label{sec:sec2}3-3-1 Models}
In the subsequent discussion, we work the electroweak gauge group $SU(3)_c\otimes SU(3)_L\otimes U(1)_X$, expanding the electroweak sector of the SM, $SU(2)_L\otimes U(1)_Y$, to $SU(3)_L\otimes U(1)_X$. Furthermore, we assume that, similar to the SM, the color group $SU(3)_c$ is vector-like (i.e., anomaly-free). Left-handed quarks (color triplets) and left-handed leptons (color singlets) transform under the two fundamental representations of $SU(3)_L$ (i.e., 3 and $3^*$).

Two categories of models will emerge: universal single-family models, where anomalies cancel within each family similar to the SM, and family models, where anomalies are canceled through interactions among multiple families.

In the context of 3-3-1 models, the most complete electric charge operator for this electroweak sector is
\begin{equation}\label{qem}
Q=\alpha\,T_{L3}+ \beta T_{L8} +X \mathbb{1},
\end{equation}
here, $T_{La}=\lambda_{a}/2$, where $\lambda_a,;a=1,2,\dots ,8$ represents the Gell-Mann matrices for $SU(3)_L$ normalized as Tr$(\lambda_a \lambda_b)=2\delta_{ab}$, and $\mathbb{1}=\text{Diag}(1,1,1)$ is the diagonal $3\times 3$ unit matrix. Assuming $\alpha=1$, the $SU(2)_L$ isospin group of the SM is fully covered in $SU(3)_L$. The parameter $\beta =\frac{2b}{\sqrt{3}}$ is a free parameter that defines the model~($\beta$ is proportional to $b$ present in the electric charge of the exotic vector boson $K_{\mu}$). The $X$ values are determined through anomaly cancellation. The 8 gauge fields $A_\mu^a$ of $SU(3)_L$ can be expressed as~\cite{Ponce:2001jn,Ponce:2002sg}
\begin{equation}\label{gfi}
\sum_a\lambda_a A^a_\mu=\sqrt{2}\left(
\begin{array}{ccc}
D^0_{1\mu} & W^+_\mu & K_\mu^{(b+1/2)} \\
W^-_\mu & D^0_{2\mu} & K_\mu^{(b-1/2)} \\
K_\mu^{-(b+1/2)} & K_\mu^{-(b-1/2)} & D^0_{3\mu} \\
\end{array}\right),
\end{equation}
here, $D^0_{1\mu}=A_\mu^3/\sqrt{2}+A_\mu^8/\sqrt{6}$, $D^0_{2\mu}=-A_\mu^3/\sqrt{2}+A_\mu^8/\sqrt{6}$, and $D^0_{3\mu}= -2A_\mu^8/\sqrt{6}$. The superscripts on the gauge bosons in Eq.~\eqref{gfi} indicate the electric charge of the particles, some of which are functions of the parameter $b$.

\subsection{The Minimal Model}
In references~\cite{Valle:1983dk,Frampton:1992wt}, it was demonstrated that, for $b=3/2$ (i.e., $\beta =\sqrt{3}$), the following fermion structure is free of all gauge anomalies: $\psi_{lL}^T= (l^-,\nu_l^0,l^+)_L\sim (1,3^*,0)$, $Q_{iL}^T=(u_i,d_i,X_i)_L\sim (3,3,-1/3)$, and $Q_{3L}^T (d_3,u_3,Y)\sim (3,3^*,2/3)$, where $l=e,\mu,\tau$ represents the {lepton family index}, $i=1,2$ for the first two quark families, and the quantum numbers after the tilde ($\sim$) denote the 3-3-1 representation. The right-handed fields are $u_{aL}^c\sim (3^*,1,-2/3)$, $d_{aL}^c\sim (3^*,1,1/3)$, $X_{iL}^c\sim(3^*,1,4/3)$, and $Y_L^c\sim (3^*,1,-5/3)$, where $a=1,2,3$ is the quark family index, and there are three exotic quarks with electric charges: $-4/3$ and 5/3. This version is referred to as \emph{minimal} in the literature because it avoids the use of exotic leptons, including possible right-handed neutrinos.


\section{\label{sec:3}Lepton and quark generations}
In what follows, we will propose sets of leptons $S_{Li}$ and quarks $S_{Qi}$  containing triplets~(anti-triplets) and singlets of $SU(3)$.  These sets must contain at least one SM generation of SM fermions. 
From Eq.~\eqref{qem}, for $\beta=\sqrt{3}$, the electric charges of the 3 and $3^*$ triplets are:  $Q_{\text{QED}}(3)= \text{Diag}(1+X,X,-1+X)$
and   $Q_{\text{QED}}(3^\star)= \text{Diag}(-1+X,X,1+X)$, respectively. 
The general expressions for the $Z'$ charges, with the $Z-Z'$ mixing angle equals to zero,  are shown in Appendix~\ref{sec:zprimecharges}.
For the SM fields embedded in the sets: $S_{L1}$, $S_{L2}$,$S_{L3}$, $S_{Q1}$  and $S_{Q2}$,  the $Z'$ charges are shown in Tables~\ref{tab:c1},\ref{tab:c2},\ref{tab:c3}, \ref{tab:c4} and  \ref{tab:c5}, respectively.
\begin{itemize}
\item 
Lepton generation  $S_{L1}=[(\nu^0_e,e^-,E_2^{--})\oplus e^+\oplus E_2^{++}]_L$ with quantum numbers $(1,3, -1);(1,1,1)$ and $(1,1,2)$ respectively. The $Z'$ charges for the SM fields are shown in Table~\ref{tab:c1}:
\begin{table}[h!]
\begin{center}
\begin{tabular}{| c | c | c |}
\hline
\multicolumn{3}{|c|}{\ \ \  $\ell=(\nu_L,e_L)^T\subset 3$, $e_R\subset 1$  (as in $S_{L1}$) }\\
\hline
fields & $g_{Z'}\epsilon_{L}^{Z'}$ & $g_{Z'}\epsilon_{R}^{Z'}$ \\ \hline
$\nu_e$ & $\frac{g_L}{\cos\theta_W} \frac{2\cos^2\theta_W-3}{\sqrt{3(1-4\sin^2\theta_W )}}$ & 0  \\ \hline
$e$     & $\frac{g_L}{\cos\theta_W} \frac{2\cos^2\theta_W-3}{\sqrt{3(1-4\sin^2\theta_W )}}$ & $\frac{g_L}{\cos\theta_W} \frac{\sqrt{3}\sin^2\theta_W}{\sqrt{1-4\sin^2\theta_W } } $ \\ \hline
\hline
\end{tabular}
\caption{$Z'$ chiral charges for the SM leptons and the right-handed neutrino when embedded  in $S_{L1}$. Here,
$\theta_W$ is the electroweak mixing angle.}
\label{tab:c1}
\end{center}
\end{table}
\item
Set $S_{L2}=[(e^-,\nu_e^0,E_1^+)\oplus e^+\oplus E_1^-]_L$ with quantum numbers $(1,3^*,0)$; $(1,1,1)$ and $(1,1,-1)$, respectively. The  $Z'$ charges for the SM fields are shown in Table~\ref{tab:c2}:
\begin{table}[h!]
\begin{center}
\begin{tabular}{| c | c | c |}
\hline
\multicolumn{3}{|c|}{\ \ \  $\ell=(\nu_L,e_L)^T\subset 3^*$, $e_R\subset 1$  (as in  $S_{L2}$) }\\
\hline
fields & $g_{Z'}\epsilon_{L}^{Z'}$ & $g_{Z'}\epsilon_{R}^{Z'}$ \\ \hline
$\nu_e$ & $\frac{g_L}{\cos\theta_W} \frac{\sqrt{1-4\sin^2\theta_W }}{2\sqrt{3}}$ & 0  \\ \hline
$e$     & $\frac{g_L}{\cos\theta_W} \frac{\sqrt{1-4\sin^2\theta_W }}{2\sqrt{3}}$ & $\frac{g_L}{\cos\theta_W} \frac{\sqrt{3}\sin^2\theta_W}{\sqrt{1-4\sin^2\theta_W } }  $ \\ \hline
\hline
\end{tabular}
\caption{$Z'$ chiral charges for the SM leptons and the right-handed neutrino when embedded  in $S_{L2}$. Here, 
$\theta_W$ is the electroweak mixing angle.}
\label{tab:c2}
\end{center}
\end{table}
\item 
Set $S_{L3}=[(e^-,\nu_e^0,e^+)]_L$ with quantum numbers $(1,3^*,0)$. The  $Z'$ charges for the SM fields are shown in Table~\ref{tab:c3}: 
\begin{table}[h!]
\begin{center}
\begin{tabular}{| c | c | c |}
\hline
\multicolumn{3}{|c|}{\ \ \  $\ell=(\nu_L,e_L)^T, e_R\subset 3^*$ (as in $S_{L3}$) }\\
\hline
fields & $g_{Z'}\epsilon_{L}^{Z'}$ & $g_{Z'}\epsilon_{R}^{Z'}$ \\ \hline
$\nu_e$ & $\frac{g_L}{\cos\theta_W} \frac{\sqrt{1-4\sin^2\theta_W }}{2\sqrt{3}}$ & 0  \\ \hline
$e$     & $\frac{g_L}{\cos\theta_W} \frac{\sqrt{1-4\sin^2\theta_W }}{2\sqrt{3}}$ & $\frac{g_L}{\cos\theta_W} \frac{\sqrt{1-4\sin^2\theta_W }}{\sqrt{3}}$ \\ 
\hline
\hline
\end{tabular}
\caption{$Z'$ chiral charges for the SM leptons and the right-handed neutrino when embedded  in $S_{L3}$. Here, 
$\theta_W$ is the electroweak mixing angle.}
\label{tab:c3}
\end{center}
\end{table}

\item 
Set $S_{Q1}=[(d,u,Q_1^{5/3})\oplus u^c\oplus d^c\oplus Q_1^{c}]_L$ with quantum numbers $(3,3^*, 2/3)$; $(3^*,1,-2/3)$; $(3^*,1,1/3)$ and $(3^*,1,-5/3)$, respectively. The $Z'$ for the SM fields are shown in Table~\ref{tab:c4}:
\begin{table}[h!]
\begin{center}
\begin{tabular}{| c | c | c |}
\hline
\multicolumn{3}{|c|}{\ \ \ $q=(u_L,d_L)^T\subset 3^*$, $u_R,d_R\subset 1$ (as in $S_{Q1}$) }\\
\hline
fields & $g_{Z'}\epsilon_{L}^{Z'}$ & $g_{Z'}\epsilon_{R}^{Z'}$ \\ \hline
$u$ 
& $ \frac{g_L}{\cos\theta_W}\frac{1}{2\sqrt{3\left(1-4\sin^2\theta\right)}}$ 
& $-\frac{g_L}{\cos\theta_W}
\frac{2\sin^2\theta_W}{\sqrt{3\left(1-4\sin^2\theta\right)}}$ \\ \hline
$d$ 
& $ \frac{g_L}{\cos\theta_W}\frac{1}{2\sqrt{3\left(1-4\sin^2\theta\right)}}$ 
& $\frac{g_L}{\cos\theta_W}
\frac{\sin^2\theta_W}{\sqrt{3\left(1-4\sin^2\theta\right)}}$ \\ \hline
\hline
\end{tabular}
\caption{$Z'$ chiral charges for the SM quarks when they are embedded  in $S_{Q1}$. Here, 
$\theta_W$ is the electroweak mixing angle.}
\label{tab:c4}
\end{center}
\end{table}

\item 
Set $S_{Q2}=[(u,d,Q_{2}^{-4/3})\oplus u^c\oplus d^c\oplus Q_{2}^c]_L$ with quantum numbers
$(3,3,-1/3)$; $(3^*,1,-2/3)$; $(3^*,1,1/3)$ and $(3^*,1,4/3)$, respectively. The $Z'$ charges for the SM fields are shown in Table~\ref{tab:c5}:

\begin{table}[h!]
\begin{center}
\begin{tabular}{| c | c | c |}
\hline
\multicolumn{3}{|c|}{\ \ \ $q=(u_L,d_L)^T\subset 3$, $u_R,d_R\subset 1$  (as in $S_{Q2}$) }\\
\hline
fields & $g_{Z'}\epsilon_{L}^{Z'}$ & $g_{Z'}\epsilon_{R}^{Z'}$ \\ \hline
$u$ &
$ \frac{g_L}{\cos\theta_W}\frac{1-2\sin^2\theta_W}{2\sqrt{3\left(1-4\sin^2\theta\right)}}$ 
& $-\frac{g_L}{\cos\theta_W}\frac{2\sin^2\theta_W}{\sqrt{3\left(1-4\sin^2\theta\right)}}$ \\ \hline
$d$ 
&  
$ \frac{g_L}{\cos\theta_W}\frac{1-2\sin^2\theta_W}{2\sqrt{3\left(1-4\sin^2\theta\right)}}$ 
& $\frac{g_L}{\cos\theta_W}
\frac{\sin^2\theta_W}{\sqrt{3\left(1-4\sin^2\theta\right)}}$ \\ \hline
\hline
\end{tabular}
\caption{$Z'$ chiral charges for the SM quarks when they are embedded  in $S_{Q2}$. Here, 
$\theta_W$ is the electroweak mixing angle.}
\label{tab:c5}
\end{center}
\end{table}
\item 
To cancel anomalies, it is advantageous introducing triplets and anti-triplets of exotic leptons; for example,  $S_{E1}=[(N_{1}^{0},E_{4}^{+},E_{3}^{++})\oplus E_{4}^{-}\oplus E_{3} ^{--}]_L$ with quantum numbers $(1,3^*,1)$; $(1,1,-1)$ and $(1,1,-2)$, respectively. We do not report the $Z'$ charges of exotic fermion fields because we assume they have a very high mass.  
\item 
Additional exotic lepton sets. $S_{E2}=[(E_{5}^{+},N_{2}^{0},E_{6}^{-})\oplus E_{5}^{-}\oplus E_{6}^{ +}]_L$ with quantum numbers $(1,3,0)$ ; $(1,1,-1)$ and $(1,1,1)$, respectively. A more economical set is 
$S_{E3}=[(E_{5}^{+},N_{2}^{0},E_{5}^{-})]$ which has  identical contributions to the anomalies as $S_{E2}$ but different  $Z'$ charges. However, these details are irrelevant for the low energy phenomenology, so we do not include $S_{E3}$ in Table~\ref{tab:anomalias}.
\end{itemize}

\section{\label{sec:4}Irreducible anomaly free sets and models}
Table~\ref{tab:anomalias} shows the contribution of each set to the anomalies. 
\begin{table}[b]
\centering
\begin{tabular}{|l|ccccccc|}\hline
Anomalías & $S_{L1}$ & $S_{L2}$ & $S_{L3}$ & $S_{Q1}$ & $S_{Q2}$ & $S_{E1}$ & $S_{E2}$ \\ \hline
$[SU(3)_C]^2U(1)_X$ &\ \ 0 &\ \ 0 &\ \ 0 &\ \ 0 &\ \ 0 &\ \ 0 &\ \ 0 \\
$[SU(3)_L]^2U(1)_X$ & $-1$ &\ \ 0 &\ \ 0 &\ \ 2 & $-1$ &\ \ 1 &\ \ 0\\
$[\text{Grav}]^2U(1)_X$    &\ \ 0 &\ \ 0 &\ \ 0 &\ \ 0 &\ \ 0 &\ \ 0 &\ \ 0\\
$[U(1)_X]^3$        &\ \ 6 &\ \ 0 &\ \ 0 &  $-12$ &\ \ 6 &\  $-6$ &\ \ 0\\
$[SU(3)_L]^3$       &\ \ 1 & $-1$ & $-1$ & $-3$ &\ \ 3 & $-1$ &\ \ 1\\
\hline
\end{tabular}
\caption{Contribution to the anomalies for each family of quarks $S_{Q_i}$, leptons $S_{L_i}$  and exotics $S_{E_i}$, for 3-3-1 models with $\beta=\sqrt{3}$.}
\label{tab:anomalias}
\end{table}
From Table~\ref{tab:anomalias}, it is possible to obtain the irreducible anomaly-free sets~\cite{Benavides:2021pqx}, shown in Table~\ref{tab:afs}.
The irreducible AFSs $Q^I_i$, $Q^{II}_i$ and  $Q^{III}_i$  in Table~\ref{tab:afs} 
correspond to fermion sets with one quark family, two quark families, or three quark families, respectively.
These sets can be combined to build three family models as
shown in Table~\ref{tab:models}. There are 33 different models (without considering all the possible embeddings). These models can also be extended by adding vector-like lepton sets, $L_i$, indicated in the second column of Table~\ref{tab:afs}.
To exemplify the possible embeddings we show some cases in Table~\ref{tabl5}.
The choice of models in Table~\ref{tabl5} show how the phenomenology depends on the SM fermion embedding in the model. For example, in the case of M10, the embedding determines whether it is strongly coupled. M17 was chosen because it had several embeddings. M3 is the minimal model. M4 is similar to the minimal model but is not universal in the lepton sector. 
\begin{widetext}
\centering
\scalebox{.9}{
\begin{tabular}{|c|l|l|l|l|}\hline
$i$& Vector-like lepton set ($L_i$)  & One quark set ($Q_i^I$)& Two quarks set ($Q_i^{I\!I}$) & Three quarks set ($Q_i^{I\!I\!I})$ \notag\\
\hline 
   1& $S_{E2}+S_{L2}$   &  $S_{E2}+2S_{L1}+S_{Q1}$           & $S_{L1}+S_{L2}+S_{Q1}+S_{Q2}$       &$3S_{L1} +2S_{Q1} +S_{Q2}$ \\
  2 & $S_{E1}+S_{L1}$   &  $S_{E1}+2S_{L2}+S_{Q2}$           & $S_{L1}+S_{L3}+S_{Q1}+S_{Q2}$       &$3S_{L2} +S_{Q1} +2S_{Q2}$\\
  3 & $S_{E2}+S_{L3}$   &  $S_{E1}+S_{L2}+S_{L3}+S_{Q2}$  &                                     &$3S_{L3} +S_{Q1} +2S_{Q2}$\\ 
  4 &                   & $S_{E1}+2S_{L3}+S_{Q2}$            &                                     &$2S_{L2} +S_{L3} +S_{Q1}+2S_{Q2}$\\
 5  &                   &                                    &                                     &$S_{L2} +2 S_{L3}+S_{Q1}+2S_{Q2}$\\
\hline
\end{tabular}}
\captionof{table}{\small AFSs for $\beta = \sqrt{3}$. We have classified the AFS according to the content of quark families, i.e.,  $Q^I_i$, $Q^{II}_i$, and $Q^{III}_i$. Combinations of these sets with three SM quark and three SM lepton families can be considered as 3-3-1 models.}
\label{tab:afs}
\end{widetext}

\begin{widetext}
\centering
\scalebox{.93}{
\begin{tabular}{|l|l|c|}\hline
&Models&\notag\\
\hline 
M1&$Q^{III}_1$& $3S_{L1} +2S_{Q1} +S_{Q2}$\\
M2&$Q^{III}_2$& $3S_{L2} +S_{Q1} +2S_{Q2}$\\
M3&$Q^{III}_3$& $3S_{L3} +S_{Q1} +2S_{Q2}$\\
M4&$Q^{III}_4$&$2S_{L2} +S_{L3} +S_{Q1}+2S_{Q2}$\\
M5&$Q^{III}_5$&$S_{L2} +2 S_{L3}+S_{Q1}+2S_{Q2}$\\
M6&$Q^{II}_1+Q^{I}_1$&$3S_{L1}+S_{L2}+S_{E2}+2S_{Q1}+S_{Q2}$ \\
M7&$Q^{II}_1+Q^{I}_2$&$S_{L1}+3S_{L2}+S_{E1}+S_{Q1}+2S_{Q2}$\\
M8&$Q^{II}_1+Q^{I}_3$&$S_{L1}+2S_{L2}+S_{L3}+S_{E1}+S_{Q1}+2S_{Q2}$\\
M9&$Q^{II}_1+Q^{I}_4$&$S_{L1}+S_{L2}+2S_{L3}+S_{E1}+S_{Q1}+2S_{Q2}$\\
M10&$Q^{II}_2+Q^{I}_1$&$3S_{L1}+S_{L3}+S_{E2}+2S_{Q1}+S_{Q2}$\\
M11&$Q^{II}_2+Q^{I}_2$&$S_{L1}+2S_{L2}+S_{L3}+S_{E1}+S_{Q1}+2S_{Q2}$\\
M12&$Q^{II}_2+Q^{I}_3$&$S_{L1}+S_{L2}+2S_{L3}+S_{E1}+S_{Q1}+2S_{Q2}$\\
M13&$Q^{II}_2+Q^{I}_4$&$S_{L1}+3S_{L3}+S_{E1}+S_{Q1}+2S_{Q2}$\\
M14&$Q^{I}_1+Q^{I}_2+Q^{I}_3$&$2S_{L1}+3S_{L2}+S_{L3}+2S_{E1}+S_{E2}+S_{Q1}+2S_{Q2}$\\
M15&$Q^{I}_1+Q^{I}_2+Q^{I}_4$&$2S_{L1}+2S_{L2}+2S_{L3}+2S_{E1}+S_{E2}+S_{Q1}+2S_{Q2}$\\
M16&$Q^{I}_1+Q^{I}_3+Q^{I}_4$&$2S_{L1}+S_{L2}+3S_{L3}+2S_{E1}+S_{E2}+S_{Q1}+2S_{Q2}$\\
M17&$Q^{I}_2+Q^{I}_3+Q^{I}_4$&$3S_{L2}+3S_{L3}+3S_{E1}+3S_{Q2}$\\
M18&$3Q^{I}_1$&$6S_{L1}+3S_{E2}+3S_{Q1}$\\
M19&$2Q^{I}_1+Q^{I}_2$&$4S_{L1}+2S_{L2}+2S_{E2}+S_{E1}+2S_{Q1}+S_{Q2}$\\
M20&$2Q^{I}_1+Q^{I}_3$&$4S_{L1}+S_{L2}+S_{L3}+S_{E1}+2S_{E2}+2S_{Q1}+S_{Q2}$\\
M21&$2Q^{I}_1+Q^{I}_4$&$4S_{L1}+2S_{L3}+S_{E1}+2S_{E2}+2S_{Q1}+S_{Q2}$\\
M22&$3Q^{I}_2$&$6S_{L2}+3S_{E1}+3S_{Q2}$ \\
M23&$2Q^{I}_2+Q^{I}_1$&$2S_{L1}+4S_{L2}+2S_{E1}+S_{E2}+S_{Q1}+2S_{Q2}$ \\
M24&$2Q^{I}_2+Q^{I}_3$&$5S_{L2}+S_{L3}+3S_{E1}+3S_{Q2}$\\
M25&$2Q^{I}_2+Q^{I}_4$&$4S_{L2}+2S_{L3}+3S_{E1}+3S_{Q2}$\\
M26&$3Q^{I}_3$&$3S_{L2}+3S_{L3}+3S_{E1}+3S_{Q2}$\\
M27&$2Q^{I}_3+Q^{I}_1$&$2S_{L1}+2S_{L2}+2S_{L3}+2S_{E1}+S_{E2}+S_{Q1}+2S_{Q2}$\\
M28&$2Q^{I}_3+Q^{I}_2$&$4S_{L2}+2S_{L3}+3S_{E1}+3S_{Q2}$\\
M29&$2Q^{I}_3+Q^{I}_4$&$2S_{L2}+4S_{L3}+3S_{E1}+3S_{Q2}$\\
M30&$3Q^{I}_4$&$6S_{L3}+3S_{E1}+3S_{Q2}$\\
M31&$2Q^{I}_4+Q^{I}_1$&$2S_{L1}+4S_{L3}+2S_{E1}+S_{E2}+S_{Q1}+2S_{Q2}$\\
M32&$2Q^{I}_4+Q^{I}_2$&$2S_{L2}+4S_{L3}+3S_{E1}+3S_{Q2}$\\
M33&$2Q^{I}_4+Q^{I}_3$&$S_{L2}+5S_{L3}+3S_{E1}+3S_{Q2}$\\
\hline
\end{tabular}}
\captionof{table}{Three-family models built from the irreducible anomaly-free sets~(Table~\ref{tab:afs}). 
It is possible to obtain (trivially) new models by adding vector-like lepton sets; we are not considering these possibilities in our counting unless they are necessary to complete the lepton families.
}
\label{tab:models}
\end{widetext}
In general, we obtain three classes of models as we can see below:
\begin{itemize}
    \item Completely non-universal models:  
    This happens if we embed each of the SM families in different sets; for example, one of the possible embeddings for the M12 model  in Table~\ref{tab:models} is to put the first lepton family in $S_{L3}$ and the remaining lepton families  in $S_{L1}$ and $S_{L2}$. This class of models usually has very strong restrictions from FCNC and CLFV. 
\item 
Universal Models:  In several AFSs,  there are embeddings with the three families of SM leptons in sets with the same quantum numbers; the same applies for the three families of the SM quarks. For example, in the M26 model in Table~\ref{tab:models}, it is possible to embed all the three SM families  in the sets $3S_{L3}+3S_{Q2}$. The remaining fields are considered exotic fermions and are necessary to cancel anomalies.
\item  The $2+1$ models:  
Most AFSs have embeddings where two families are in sets with the same quantum numbers, and the  third {family is a different} set. To avoid the strongest FCNC restrictions,  it is necessary that the left-handed doublets of the first two SM quark families have identical quantum numbers.  This condition is also desirable for Lepton families, although some models could avoid the FCNC constraints without satisfying this condition. A typical example of these models is
the Pisano-Pleitez-Frampton minimal model~\cite{Pisano:1992bxx,Frampton:1992wt}.
$3S_{L3} +S_{Q1} +2S_{Q2}$ (the M3 model in Table~\ref{tab:models}). This model is universal in the lepton sector and non-universal in the quark sector.
\end{itemize}



\section{\label{sec:5}LHC and low energy constraints}
We consider the ATLAS search for high-mass dilepton resonances in the mass range of 250~GeV to 6~TeV in proton-proton collisions at a center-of-mass energy of $\sqrt{s}=13$ TeV during Run 2 of the LHC with an integrated luminosity of 139~fb$^{-1}$~\cite{ATLAS:2019erb}. This data was collected from searches of $Z^{\prime}$ bosons decaying dileptons.
We obtain the lower limit on the $Z'$ mass from the intersection of the theoretical predictions for the cross-section with the corresponding  upper limit reported by ATLAS at a 95\% confidence level.   We use the expressions given in  Ref.~\cite{Erler:2011ud,Salazar:2015gxa,Benavides:2018fzm} to calculate the theoretical cross-section. We assume that the $Z-Z'$ mixing angle $\theta$~(see appendix~\ref{sec:zzmixing}) equals zero for these bounds.
\begin{table}[b!]
\begin{center}
\begin{tabular}{|c|c|}
\hline
       Particle content                            &    LHC-Lower limit   \\  
       first generation                            &     in TeV           \\   \hline 
$S_{L3}+S_{Q1} $ &  7.3    \\ \hline
$S_{L3}+S_{Q2} $ &  6.4    \\ \hline 
\end{tabular}
\caption{
The lepton families $S_{L_1}$  and $S_{L_2}$ are strongly coupled (For $S_{L_1}$ and  $S_{L_2}$ the left-handed lepton doublet $\ell$ and  the right-handed  charged lepton singlet $e_R$   have couplings greater than 1, respectively). Therefore only $S_{L_3}$ is phenomenologically viable for the first family. Depending on the quark content, i.e.,  $S_{Q_1}$ or $S_{Q_2}$,  we have two different constraints. }
\label{tabla8}
\end{center}
\end{table}
In Table~\ref{tabla8}, the LHC constraints for some models are presented. It is important to stress that the leptons of the first family, i.e.,  the electron and its neutrino,  should be embedded in $S_{L3}$ since it is the only scenario where the right-handed electron has $Z'$ couplings less than 1.  
In Table~\ref{tabla8}, this is the best option for models with the first two lepton generations embedded in $S_{L3}$,  as it happens for the minimal model~(M3), since having identical quantum numbers for the first and second lepton families avoids possible issues with CLFV and FCNC. 
 To avoid the strongest FCNC constraints in the quark sector, the charges of the left-handed quarks of the first two families should be identical~\cite{Langacker:2008yv}; this feature is assumed to calculate the lower mass limits in Table~\ref{tabla8}.
It is important to stress the non-universal $Z'$ couplings modify    
 processes such as~\cite{Langacker:2000ju}: coherent $\mu-e$ conversion in a muon atom, $K^0-\Bar{K}^0$ and $B-\Bar{B}$ mixing, $\epsilon$, and $\epsilon'/\epsilon$, 
lepton, and semileptonic decays (e.g.,  $\mu\to e\gamma$) which, 
if observed in the future, the Non-Universal Models will be favored over the Universal ones.
For models with a $Z'$ boson coupling in a different way to the third family, there are different predictions for the branching rations  $B(t\rightarrow Hu)$ and $B(t\rightarrow Hc)$. These predictions are strongly constrained by colliders~\cite{CMS:2021gfa}. 
 In Table~\ref{tabl5}, SC stands for strongly coupled, indicating that in the sets $S_{L1}$ and $S_{L2}$, the coupling of the right-handed electron is greater than one, and therefore, the collider constraints are very strong.
Even though $Z'$ with couplings greater than one to the SM fields of the first generation are quite disfavored by colliders~\cite{Erler:2011ud},  strongly coupled models are also attractive in several phenomenological approaches~\cite{Langacker:2008yv,Hill:2002ap}; for this reason, it is important to realize the existence of these models, which naturally appear  in 3-3-1 models with large $\beta$ values. 
Regarding constraints on exotic particles, the restrictions on the mass of a sequential heavy lepton are above 100 GeV~\cite{ParticleDataGroup:2022pth}. For exotic quarks $t'$ and $b'$, the allowed mass ranges  are above 1370~GeV and 1570~GeV, respectively~\cite{ParticleDataGroup:2022pth}. The restrictions on fields with exotic electric charges are weaker because the identification algorithms assume the charges are proportional to the charge of the electron~\cite{RomeroAbad:2020uvo}.
The presence of doubly charged exotic leptons can generate new decay channels in proton-proton collisions at very high energies. In  Figure~\ref{fig1}, the Feynman diagram for the process
$q\bar{q}\rightarrow Z'\rightarrow E^{++}E^{--}\rightarrow \ell^{+}\ell^{-}\gamma \rightarrow \ell^{+}\ell^{-}\mu^{+}\mu^{-}(\tau^{+}\tau^{-}) $, generating four boosted leptons in the final state (the doubly charged exotic lepton appears in $S_{L1}$, which strongly couples the $Z'$; for this reason, to avoid collider constraints, we restrict to leptons of the second or third family). 
On the other hand, exotic quarks modify the $K^{0}-\bar{K}^{0}$ mixing, as shown in Figure~\ref{fig2}.  Fermions with exotic electric charges can contribute to several processes; however, an exhaustive study of these processes is beyond the purpose of this work.
%
\begin{widetext}
\begin{center}
\centering 
\begin{figure}
\scalebox{0.85}{
\begin{tikzpicture}
\begin{feynman}
  \vertex (a);
  \vertex [above left=of a] (b) {$\bar q$};
   \vertex [below left=of a] (c) {$q$};
   \vertex[right=of a] (d);
   \vertex[above right=of d] (e);
   \vertex[above right=of e] (e1) {$\mu^+,\tau^+$};
   \vertex[below right=of d] (f);
   \vertex[below right=of f] (f1){$\mu^-,\tau^-$};
   \vertex[above right=of f] (f2);
   \vertex[right=of f2] (f3);
    \vertex[above right=of f3] (g1){$\ell^+$};
    \vertex[below right=of f3] (g2){$\ell^-$};
   
   \diagram{
   (c) -- [fermion]  (a) -- [fermion] 
   (b);
   (a) -- [boson, edge label=$Z'$] (d);
   (d) -- [fermion, edge  label=$E^{++}$](e);
    (f) -- [fermion, edge  label=$E^{--}$](d);
     (e) -- [fermion]  (e1);
      (f1) -- [fermion]  (f);
      (e) -- [boson, edge label=$K^+$] (f2);
      (f) -- [boson, edge label=$K^-$] (f2);
       (f2) -- [boson,edge label=$\gamma$] (f3);
       (g1) -- [fermion] (f3);
       (f3) -- [fermion] (g2);
  };
   \end{feynman}
\end{tikzpicture}
}
\caption{Doubly charged exotic lepton contribution  to the  process 
$q\bar{q}\rightarrow Z'\rightarrow E^{++}E^{--}\rightarrow \ell^{+}\ell^{-}\gamma \rightarrow \ell^{+}\ell^{-}\mu^{+}\mu^{-}(\tau^{+}\tau^{-}) $.}\label{fig1}
\end{figure}
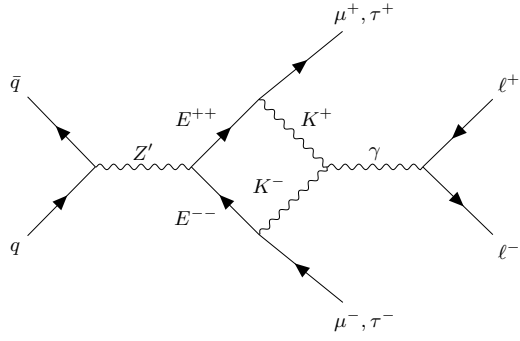
\end{center}


\begin{center}
\centering 
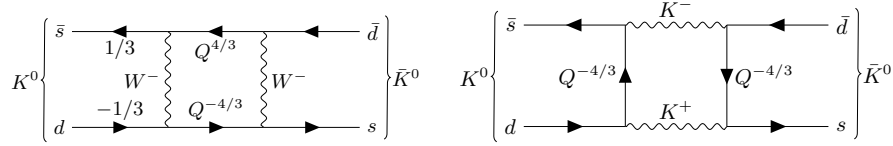
\begin{figure}
\begin{tabular}{cc}
\scalebox{0.85}{
\begin{tikzpicture}
 \begin{feynman}
  \vertex (a);
  \vertex [below=of a] (b);
  \vertex [left=of a] (c) {$\bar s$};
  \vertex [left=of b] (d) {$ d$};
  \vertex [right=of a] (e);
  \vertex [right=of b] (f);
  \vertex [right=of e] (g) {$\bar d$};
  \vertex [right=of f] (h) {$s$};

   \diagram{
   (a) -- [fermion, edge label=$1/3$]  (c); 
   (d) -- [fermion, edge label=$-1/3$]  (b); 
   (b) -- [boson, edge label=$W^-$]  (a); 
   (e) -- [fermion, edge label=$Q^{4/3}$] (a);
   (b) -- [fermion, edge label=$Q^{-4/3}$] (f);
   (e) -- [boson, edge label=$W^{-}$]  (f); 
   (g) -- [fermion]  (e); 
    (f) -- [fermion]  (h); 
  };
  \draw [decoration={brace}, decorate] (d.south west) -- (c.north west)
node [pos=0.5, left] {\(K^{0}\)};
 \draw [decoration={brace}, decorate] (g.north east) -- (h.south east)
node [pos=0.5, right] {\(\bar K^{0}\)};
   \end{feynman}
\end{tikzpicture}
}
&
\scalebox{0.9}{
\begin{tikzpicture}
 \begin{feynman}
  \vertex (a);
  \vertex [below=of a] (b);
  \vertex [left=of a] (c) {$\bar s$};
  \vertex [left=of b] (d) {$ d$};
  \vertex [right=of a] (e);
  \vertex [right=of b] (f);
  \vertex [right=of e] (g) {$\bar d$};
  \vertex [right=of f] (h) {$s$};

   \diagram{
   (a) -- [fermion]  (c); 
   (d) -- [fermion]  (b); 
   (b) -- [fermion, edge label=$Q^{-4/3}$]  (a); 
   (a) -- [boson, edge label=$K^{-}$] (e);
   (b) -- [boson, edge label=$K^{+}$] (f);
   (e) -- [fermion, edge label=$Q^{-4/3}$]  (f); 
   (g) -- [fermion]  (e); 
    (f) -- [fermion]  (h); 
  };
  \draw [decoration={brace}, decorate] (d.south west) -- (c.north west)
node [pos=0.5, left] {\(K^{0}\)};
 \draw [decoration={brace}, decorate] (g.north east) -- (h.south east)
node [pos=0.5, right] {\(\bar K^{0}\)};
   \end{feynman}
\end{tikzpicture}
}
\end{tabular}
\caption{Exotic quark contribution  to the $K^{0}-\bar{K}^{0}$ mixing.}\label{fig2}
\end{figure}
\end{center}
%
\begin{center}
\scalebox{0.84}{
\begin{tabular}{|l|l|lcccc|}\hline
&&&&&&\\[-3.5mm]
{ Model}&$j$&{ SM Lepton Embeddings}&{ Universal} |&$2+1$ & | { Quark Configuration}& | { LHC-Lower limit}
\\
&&&&&&\\[-3.5mm]
\hline
&&&&&&\\[-3.5mm]
\multirow{1}{*}{$M3=Q_3^{I\!I\!I}$(Minimal)}&-& [{ $3S_{L3}^{\bar\ell+e^{\prime+}}$}]&\checkmark&$\times$&$2S_{Q2}+S_{Q1}$
&6.4 TeV\\
&&&&&&\\[-3.5mm]\hline
&&&&&&\\[-3.5mm]
\multirow{1}{*}{$M4=Q_4^{I\!I\!I}$}&-&  [{$2S_{L2}^{\bar\ell+e^{+}}+S_{L3}^{\bar\ell+e^{\prime+}}$}]&$\times$&\checkmark&$2S_{Q2}+S_{Q1}$
&6.4 TeV\\
&&&&&&\\[-3.5mm]\hline
&&&&&&\\[-3.5mm]
\multirow{2}{*}{$M6=(Q_1^{I}+Q_1^{I\!I})^j$}&1&  [{$3S_{L1}^{\ell+e^{+}}$}]$+S_{L2}+S_{E2}$&\checkmark&$\times$&$2S_{Q1}+S_{Q2}$
& SC\\
&2&[{$2S_{L1}^{\ell+e^+}+S_{L2}^{\bar\ell+e^{+}}$}$]+S_{L1}+S_{E2}$&$\times$&\checkmark& $2S_{Q1}+S_{Q2}$
& SC\\
&&&&&&\\[-3.5mm]\hline
&&&&&&\\[-3.5mm]
\multirow{4}{*}{$M17=(Q_2^{I}+Q_3^{I}+Q_4^{I})^j$}&1&  [{$3S_{L2}^{\bar\ell+e^{+}}$}]$+3S_{L3}+3S_{E1}$&\checkmark&$\times$& $3S_{Q2}$
&SC\\
&&&&&&\\[-3.5mm]
&2&[{$3S_{L3}^{\bar\ell+e^{\prime+}}$}]$+3S_{L2}+3S_{E1}$&\checkmark&$\times$& $3S_{Q2}$
& 6.4 TeV\\
&&&&&&\\[-3.5mm]
&3&[{$2S_{L2}^{\bar\ell+e^+}+S_{L3}^{\bar\ell+e^{\prime+}}$}$]+S_{L2}+2S_{L3}+3S_{E1}$&$\times$&\checkmark& $3S_{Q2}$
& 6.4 TeV\\
&&&&&&\\[-3.5mm]
&4&[{$S_{L2}^{\bar\ell+e^+}+2S_{L3}^{\bar\ell+e^{\prime+}}$}]$+2S_{L2}+S_{L3}+3S_{E1}$&$\times$&\checkmark& $3S_{Q2}$
& 6.4 TeV\\
&&&&&&\\[-3.5mm]\hline
&&&&&&\\[-3.5mm]
\multirow{2}{*}{$M10=(Q_1^{I}+Q_2^{I\!I})^j$}&1&  [{$3S_{L1}^{\ell+e^{+}}$}$]+S_{L3}+S_{E2}$ &\checkmark&$\times$&$2S_{Q1}+S_{Q2}$
& SC \\
&2&[{$2S_{L1}^{\ell+e^+}+S_{L3}^{\bar\ell+e^{\prime+}}$}]$+S_{L1}+S_{E2}$&$\times$&\checkmark& $2S_{Q1}+S_{Q2}$
& 7.3 TeV\\
\hline
\end{tabular}
}
\captionof{table}{\small
Alternative embeddings of the SM fields for some of the models in Table~\ref{tab:models}. 
The lepton sets in square brackets contain the standard model fields. 
The superscripts correspond to the particle content of the SM, where $\ell$~($\bar{\ell}$) 
stands for  a left-handed lepton doublet embedded in a $SU(3)_L$ triplet~(anti-triplet), 
and $e^{\prime +}$~($e^+$) is the right-handed charged lepton embedded in a $SU(3)_L$ triplet~(singlet). 
The check mark $\checkmark$ means that at least two (2+1) or three (universal)  families have the same charges under the gauge symmetry. The cross $\times$ stands for the opposite.
LHC constraints are obtained  from Table~\ref{tabla8} for embeddings in which we can choose the same $Z'$ charges for the first two families,
otherwise, we leave the space blank.
To avoid a strongly coupled model in the Lepton sector, it is necessary to embed the first Lepton family (electron and electron neutrino)  in $S_{L3}$. This feature will be helpful to distinguish between the different embeddings. The embedding also defines the content of exotic particles in each case.
}  
\label{tabl5}
\end{center}
\end{widetext}
\section{Conclusions}
Since that for 3-3-1 models, the absolute value of the parameter $\beta$  must be less than $\beta\lessapprox \cot \theta_W =1.8$
(for $\sin^2\theta_W=0.231$ in the $\overline{\text{MS}}$ renormalization scheme at the $Z$-pole energy scale),
and the values of $\beta$ are further limited by the requirement that the vector boson charges be integers, the possible values of this parameter are reduced to a few cases. For a realistic model, the maximum possible value corresponds to $\beta=\sqrt{3}\sim 1.73$. This case is   important since it contains the Pleitez-Frampton minimal model. 
We have constructed three sets of lepton families, $S_{Li}$, two quark families, $S_{Qi}$, and two exotic lepton families $S_{Ei}$, and we calculated their contribution to anomalies.
In our analysis, we obtained 14 irreducible AFSs, from which we built  33 non-trivial 3-3-1 models (without considering the different embeddings) with at least three quark and three lepton families for each case.  Each of these embeddings constitutes a phenomenologically distinguishable model; however, we limited our analysis of the possible embeddings to a few cases.
In the same way, from our analysis of the 3-3-1 models with $\beta=\sqrt{3}$ 
we report  the couplings of the SM fields to the $Z'$  boson  for all the possible quark and lepton families
and the corresponding lower limits on the $Z'$ mass. 
We also discuss the conditions under which the reported models avoid FCNC and CLFV.  We also observed that strongly coupled models appear naturally and require a high value for the $Z'$ mass. They can be helpful in specific phenomenological approaches based on models with strong dynamics. In the future, a detailed analysis of each model will be necessary; however, this is beyond the scope of the present work.

\begin{acknowledgments}
This research was partly supported by the ``Vicerrectoría de Investigaciones e Interacción Social VIIS de la Universidad de Nariño'',  project numbers 2686, 1928, 2172,  2693, and 2679.
E.R, R. H. B. and Y.G. acknowledge additional financial support from Minciencias CD 82315 CT ICETEX 2021-1080.
\end{acknowledgments}
\appendix

\section{$Z'$ charges for a general 3-3-1 model\label{sec:zprimecharges}}
At low energy, the 3-3-1 models, i.e., the gauge symmetry 
$SU(3)_C \otimes SU(3)_L\otimes U(1)_X$ reduces to the low energy effective theory 
$SU(3)_C \otimes SU(2)_L\otimes U(1)_{8L}\otimes U(1)_X\rightarrow SU(3)_C \otimes SU(2)_L\otimes U(1)_Y$.
From the covariant derivatives, for the neutral currents, we obtain the interaction Lagrangian
\begin{align}
-\mathcal{L}\supset  g_L J_{3L}^{\mu}A_{3L\mu}+g_L J_{8L}^{\mu}A_{8L\mu}+g_{X} J_{X}^{\mu}A_{X\mu}\,, 
\end{align}
which can be written as 
\begin{align}\label{eq:lagrangian2}
-\mathcal{L}_{NC}=& g_{i}J_{i\mu}A^{\mu}_{i}
= g_{j}J_{j\mu}O_{jk} O_{kl}^TA^{\mu}_{l}, 
\notag\\
=&\tilde g_{k}\tilde J_{k\mu}\tilde A^{\mu}_{k}\,,
\end{align}
where $\tilde A^{\mu}_{k}=O_{kl}^TA^{\mu}_{l}$, then  $(A_1^{\mu},A_2^{\mu})=(A_{8L}^{\mu},A_{X}^{\mu})$,  $(\tilde A_1^{\mu},\tilde A_2^{\mu})=( B^{\mu},Z^{\prime\mu})$, 
$(g_1 J_1^{\mu},g_2 J_2^{\mu})=(g_L A_{8L}^{\mu}, g_X A_{X}^{\mu})$ and  $(\tilde g_1\tilde A_1^{\mu},\tilde g_2\tilde A_2^{\mu})=( g_YJ_{Y}^{\mu},g_{Z^{\prime}}J_{Z^{\prime}}^{\mu})
$.  At high energies, the symmetry is broken following the breaking chain $SU(3)_C\otimes SU(3)_L \otimes U_X(1)\rightarrow SU(3)_C\otimes SU(2)_L\otimes U_{8L}(1)\otimes U_X(1)=SU(3)_C\otimes SU(2)_L\otimes U_{Y}(1)\otimes U'(1)$, i.e.,
\begin{align}
\begin{pmatrix}
A_{3L}\\
 B^{\mu}\\
 Z^{\prime\mu}
\end{pmatrix}
=
\begin{pmatrix}
1& 0_{1\times 2}\\
0_{2\times 1}& O^{T}_{2\times 2}
\end{pmatrix}
\begin{pmatrix}
A_{3L}\\
A_{8L}^{\mu}\\
A_{X}^{\mu}
\end{pmatrix}\ .
\end{align}
Next step $SU(3)_C\otimes SU(2)_L\otimes U_{Y}(1)\otimes U'(1)\rightarrow SU(3)_C\otimes U_{\text{QED}}(1)$, i.e.,
\begin{align}
\begin{pmatrix}
A^{\mu}\\
Z^{\mu}\\
Z^{\prime\mu}
\end{pmatrix}
=
\begin{pmatrix}
 \sin\theta_W & \ \cos\theta_W& 0\\ 
 \cos\theta_W & -\sin\theta_W& 0\\
 0            &   0          & 1
\end{pmatrix}
\begin{pmatrix}
A_{3L}\\
 B^{\mu}\\
 Z^{\prime\mu}
\end{pmatrix}\,.
\end{align}
Where the fields correspond to the SM photon $A^\mu$ and the $Z^{\mu}$ boson, and a heavy vector-boson $Z^{\prime}$.
Proceeding similarly for the currents, 
and limiting ourselves to the fields on which the orthogonal submatrix $Q_{2\times 2}$ acts, 
from Eq.~\eqref{eq:lagrangian2} 
we obtain $\tilde g_k\tilde J_{k}^{\mu}= g_{j}J_{j}^{\mu}O_{jk}$,
i.e.,
\begin{align}
\tilde g_{k}\tilde J_{k\mu}= 
\begin{pmatrix}
g_Y J_{Y}^{\mu}, &  g_{Z'} J_{Z'}^{\mu}
\end{pmatrix}    
=
\begin{pmatrix}
g_{L} J_{L8}^{\mu}, &  g_{X} J_{X}^{\mu}
\end{pmatrix}
\begin{pmatrix}
O_{11}& O_{12}\\
O_{21}& O_{22}
\end{pmatrix},\notag\\ 
=
\begin{pmatrix}
g_{L} J_{L8}^{\mu} O_{11}+ g_{X} J_{X}^{\mu} O_{21},&
g_{L} J_{L8}^{\mu} O_{12}+ g_{X} J_{X}^{\mu}O_{22}
\end{pmatrix}\,.
\end{align}
Without further assumption 
\begin{align}
\begin{pmatrix}
O_{11}& O_{12}\\
O_{21}& O_{22}
\end{pmatrix}
=
\begin{pmatrix}
\cos\omega &  -\sin\omega\\
\sin\omega &   \cos\omega 
\end{pmatrix}\,, 
\end{align}
so that
\begin{align}\label{eq:jy}
g_Y J_{Y}^{\mu}     &= \ \ g_{L} J_{L8}^{\mu}\cos\omega+ g_X J_{X}^{\mu}\sin\omega,\notag\\
g_{Z'} J_{Z'}^{\mu} &=  -g_{L} J_{L8}^{\mu}\sin\omega+ g_X J_{X}^{\mu}\cos\omega\ .
\end{align}
The charge operator in a 3-dimensional representation is given by
\begin{align}
Q_{\text{QED}}=T_{L3}+\beta  T_{L8} +X\mathbb{1}\,,     
\end{align}
hence
\begin{align}
Y= \beta  T_{L8}+  X\,.
\end{align}
From this expression, it is possible to obtain a relation between the currents (the 
currents are proportional to the charges)
\begin{align}
J_Y^{\mu}= \beta  J_{L8}^{\mu}+ J_{X}^{\mu}\,.
\end{align}
Comparing this result with~\eqref{eq:jy}
\begin{align}
\beta  =\frac{g_{L} \cos\omega }{g_Y},\hspace{1cm} 1 =\frac{g_{X} \sin \omega }{g_{Y}}\,.    
\end{align}
From $\cos^2\omega+\sin^2\omega=1$, we obtain
\begin{align}
\left(\frac{\beta }{g_{L}}\right)^2+\left(\frac{1}{g_X}\right)^2 =\frac{1}{g_Y^2}\,.     
\end{align}
In the SM, $g_L\approx 0.652$ and  $g_Y=g_L\tan\theta_W$,
\begin{align}
g_X= \frac{g_L\tan\theta_W}
{\sqrt{1-\beta ^2\tan^2\theta_W}}\,.
\end{align}
This expression shows that  the  parameter $\beta$ cannot be arbitrarily large from the matching conditions $\beta\lessapprox \cot \theta_W $; some care must be taken on this approximation since this is a renormalization-scheme dependent inequality. 
From these expressions, we obtain 
\begin{align}
\cos\omega = \frac{\beta }{g_{L}}g_Y= \beta  \tan\theta_W\,,\hspace{.3cm} \sin\omega= \sqrt{1-\beta ^2 \tan^2\theta_W}\,.     
\end{align}
From Eq.~\eqref{eq:jy}, $g_{Z'} \epsilon_{Z'}=  -g_L T_{8L}\sin\omega+ g_X X_{X}\cos\omega$, we obtain
\begin{align}
g_{Z'} \epsilon_{Z'}&=
-g_L T_{L8}\sqrt{1-\beta ^2 \tan^2\theta_W}
+  \beta \frac{g_L\tan^2\theta_W X}{\sqrt{1-\beta ^2\tan^2\theta_W}},\notag\\
&=
g_L\left(-T_{L8}\tilde \alpha
+   \beta \frac{\tan^2\theta_W}{\tilde \alpha}X\right)\,,\notag\\
\end{align}
where $\tilde\alpha=\sqrt{1-\beta ^2\tan^2\theta_W}=\frac{1}{\cos\theta_W}\sqrt{1-4 \sin^2\theta_W}$ for $\beta=\sqrt{3}$.

\section{Chiral charges for the $3$ representation}
In what follows, we propose sets of fermions representing the particle content of a generation of leptons or quarks, for 
left-handed  triplets $3$, and for right-handed fermions 
in an $SU(3)_L$ singlet, in general we have 
\begin{widetext}
\begin{align}
g_{Z'}\epsilon^{Z'}_L(3)=g_L\left(\begin{array}{ccc}
-\frac{1}{2\sqrt{3}}{\tilde \alpha} 
+ \beta \frac{ \tan^2\theta_W}{\tilde \alpha}X & 0 & 0 \\
0 & -\frac{1}{2\sqrt{3}}{\tilde \alpha} +\beta \frac{ \tan^2\theta_W}{\tilde \alpha}X & 0 \\
0 & 0 & \frac{1}{\sqrt{3}}{\tilde \alpha}+\beta \frac{\tan^2\theta_W}{\tilde \alpha}X \\
\end{array}\right)\,,
\hspace{0.5cm}
\epsilon^{Z'}_{R}
= g_L\beta \frac{\tan^2\theta_W}{\tilde \alpha}X_R\, .
\end{align}
\end{widetext}
Here we add the subindex $R$ to the X-charge of the right-handed singlet to emphasize that it differs from
the quantum number of the left-handed triplet, i.e., $X$.    
If the charge conjugate of the  right-handed  fermion is identified with the third component of a triplet, then 
$\epsilon^{Z'}_R=-g_L\left(\frac{1}{\sqrt{3}}{\tilde \alpha}+\beta \frac{\tan^2\theta_W}{\tilde \alpha}X\right)$.

\section{The conjugate representation $3^\star$}
To cancel the anomalies of $SU(3)_L$, triplets must be put in the conjugate representation. In general, for any set of generators $T^{a}$ of an $SU(N)$ symmetry  with $N\le 3$ there exists another set of generators $-T^{a*}$, which satisfy the same algebra. This set of generators spawns the so-called conjugate representation of $SU(N)$. With these generators, we can build charge operators and multiplets containing the SM particles.
To compare with the conjugate representation, we use the projectors 
\begin{align}
p_{12}
=
\begin{pmatrix}
1  & 0 & 0\\
0  & 1 & 0 \\
0  & 0 & 0 
\end{pmatrix}
\,,
\hspace{0.5cm}
\tilde p_{12}
=
\begin{pmatrix}
0 &  1 & 0\\
1  & 0 & 0 \\
0  & 0 & 0 
\end{pmatrix}.
\end{align}
They should not be confused with permutation operators, as the purpose of these operators is to compare only the first two rows of the charge operators.
 $\tilde p_{12}$ also permutes the first two eigenvalues to make a proper comparison with the conjugate operator. We can obtain the $X^C$, i.e., the charge of the triplet $3^\star$ in the conjugate representation, from the equation
\begin{align}
&\tilde p_{12}\left(T_{L3}+\beta  T_{L8} +X\mathbb{1}\right)\tilde p_{12}^T\notag\\
= 
& p_{12}\left(-T_{L3}-\beta  T_{L8}+X^{C}\mathbb{1}\right) p_{12}^T\, ,
\end{align}
only the signs of the $SU(3)$ generators were changed.
This matrix equation is equivalent to a couple of linear equations. These 
equations have the solution 
$X^C=\left(\frac{\beta}{\sqrt{3}}+X\right)=\left(1+X\right)$. An equivalent treatment is to obtain the conjugate representation from $T_{3L}-\beta   T_{8L}+X^c$, which generates the exact electric charges but in a different order. We verify that both approaches contribute identically to the anomalies, developing the same particle content  and models.
For left-handed  triplets in the conjugate representation $3^\star$, and  right-handed fermions 
in an $SU(3)_L$ singlet, we have, in general,
\begin{widetext}
\begin{align}
g_{Z'}\epsilon^{Z'}_L(3^\star)=g_L\left(\begin{array}{ccc}
+\frac{1}{2\sqrt{3}}{\tilde \alpha} 
+ \beta \frac{ \tan^2\theta_W}{\tilde \alpha}X^C & 0 & 0 \\
0 & +\frac{1}{2\sqrt{3}}{\tilde \alpha} +\beta \frac{ \tan^2\theta_W}{\tilde \alpha}X^C & 0 \\
0 & 0 & -\frac{1}{\sqrt{3}}{\tilde \alpha}+\beta \frac{\tan^2\theta_W}{\tilde \alpha}X^C \\
\end{array}\right)\,,
\hspace{0.5cm}
\epsilon^{Z'}_{R}
= g_L\beta \frac{\tan^2\theta_W}{\tilde \alpha}X^C\, .
\end{align}
\end{widetext}
If the charge conjugate of the  right-handed  fermion is identified with the third component of a triplet, then 
$\epsilon^{Z'}_R=-g_L\left(-\frac{1}{\sqrt{3}}{\tilde \alpha}+\beta \frac{\tan^2\theta_W}{\tilde \alpha}X^C\right)$.

\section{$Z$-$Z'$ Mixing}
\label{sec:zzmixing}
Mixing angle $\theta$ between $Z$  and  $Z'$ is tightly constrained~\cite{Erler:2009jh}, i.e., $\theta< 10^{-3}$; however,  in several phenomenological analyses, it is still  useful delivering expressions for the mass eigenstates. 
\begin{align}
Z^{\mu}_1=&\ \ \ \ Z^{\mu}\cos\theta + Z^{\prime\mu}\sin\theta,\notag\\
Z^{\mu}_2=&-Z^{\mu}\sin\theta + Z^{\prime\mu}\cos \theta\ .\notag\\
\end{align}
At low energies, $Z_1$ is identified with the SM $Z$ boson.
In order to keep the Lagrangian invariant, this field rotation must be compensated by  the corresponding rotation of the currents,
\begin{align}
g_1J^{\mu}_1=&\ \ \ \ g_ZJ^{\mu}_Z\cos\theta + g_{Z'}J_{Z^{\prime}}^{\mu}\sin\theta,\notag\\
g_2J^{\mu}_2=&-g_ZJ_{Z}^{\mu}\sin\theta      + g_{Z'}J_{Z^{\prime}}^{\mu}\cos \theta\ .\notag\\
\end{align}
From which we get 
\begin{align}
g_1Q_1=&\ \ \ \ g_ZQ_Z\cos\theta    + g_{Z'}Q_{Z^{\prime}}\sin\theta,\notag\\
g_2Q_2=&        -g_ZQ_{Z}\sin\theta  + g_{Z'}Q_{Z^{\prime}}\cos \theta\ .\notag\\
\end{align}




\bibliography{apssamp}

\end{document}